\documentclass[twocolumn,aps,citeautoscript,epsfig,superscriptaddress,pop]{revtex4}
\usepackage{amsmath}
\usepackage{amsfonts}
\usepackage{amssymb}
\usepackage{graphicx}
\usepackage{dcolumn}
\usepackage{bm}
\usepackage[sort&compress]{natbib}
\bibpunct{[}{]}{,}{n}{,}{,}
\newcommand{\dif}[2]{\ensuremath{\frac{\partial #1}{\partial #2}}}

\begin{document}
\preprint{APS/123-QED}

\title{Plasma wake inhibition at the collision of two laser pulses in an underdense plasma}

\author{C. Rechatin}
 
\author{J. Faure}
\author{A. Lifschitz}%
\author{V. Malka}
\affiliation{%
Laboratoire d'Optique Appliqu\'ee, ENSTA, CNRS, Ecole Polytechnique, UMR 7639, 91761 Palaiseau, France\\
}%
\author{E. Lefebvre}
\affiliation{%
D\'epartement de Physique Th\'eorique et Appliqu\'ee, CEA, DAM Ile-de-France, BP 12, 91680 Bruy\`eres-le-Ch\^atel, France\
}

\date{\today}

\begin{abstract}
An electron injector concept for laser-plasma accelerator was developed in ref \cite{Esarey} and \cite{Fubiani} ; it relies on the use of
counter-propagating ultrashort laser pulses. In \cite{Fubiani}, the scheme is
as follows: the pump laser pulse generates a large amplitude laser
wakefield (plasma wave). The counter-propagating injection pulse
interferes with the pump laser pulse to generate a beatwave
pattern. The ponderomotive force of the beatwave is able to inject
plasma electrons into the wakefield. We have studied this
injection scheme using 1D Particle in Cell (PIC) simulations.
The simulations reveal phenomena and important physical processes
that were not taken into account in previous models. In
particular, at the collision of the laser pulses, most plasma
electrons are trapped in the beatwave pattern and cannot
contribute to the collective oscillation supporting the plasma wave. At this
point, the fluid approximation fails and the plasma wake is
strongly inhibited. Consequently, the injected charge is reduced by one order of magnitude compared to the predictions from previous models.
\end{abstract}

\maketitle An intense laser pulse can drive an electrostatic
plasma wave via the ponderomotive force which scales as $\nabla
{\bf a}^2$, where ${\bf a}$  is the normalized potential vector of the laser :
$|{\bf a}|=a=8.6\times10^{-10} \lambda\mbox{[$\mu$m]}
I_0^{1/2}\mbox{[W.cm$^{-2}$]}$ for a linearly polarized laser. When the pulse duration is close
to the plasma period ($\lambda_p/c$), the laser pulse
ponderomotive force pushes electrons and efficiently creates
charge separation (ions hardly move). It results in a travelling
longitudinal wave whose phase velocity $v_p$ is equal to the group
velocity of the laser. In an underdense plasma, $v_p$ is very close
to $c$, the speed of light, thus enabling acceleration of electrons
to very high energies once they are trapped in the wake \cite{Tajima}. But in a linear or moderately
nonlinear regime, an electron with no initial velocity is not
trapped by this travelling wave and consequently, not accelerated.
In a more nonlinear regime, transverse wave breaking effects can
result in the self-trapping of electrons in the so-called "bubble
regime" \cite{Bubble1}. This phenomenon has been observed in 2004
in \cite{Nature1,Nature2,Nature3} where quasi monoenergetic
electron beams at the 100 MeV level were obtained.  Nevertheless,
in this scheme, self-injection and acceleration depend on the precise evolution of the laser pulse. Therefore, a fine
control over the output electron beam is hard to achieve. On the
contrary, precise control of electron injection would translate into good tailoring of the electron beam parameters, and would be most useful for
applications \cite{DesRosiers,Yannick}.

To trap electrons in a regime where self-trapping does not occur,
one has to externally inject electrons in the plasma wave, i.e.
give electrons an initial momentum. In addition, electrons should
be injected in a short time ($<\lambda_p/c$) in order to produce a
monoenergetic beam. This can be achieved using additional
ultrashort laser pulses whose purpose is only restricted to
triggering electron injection. Umstadter et al. \cite{Umstadter}
first proposed to use a second laser pulse propagating
perpendicular to the pump laser pulse. The idea was to use the
radial ponderomotive kick of the second pulse to inject electrons.
Esarey et al. \cite{Esarey} proposed a counter-propagating
geometry based on the use of three laser pulses. This idea was
further developed in Ref. \cite{Fubiani}, where only two laser
pulses are necessary. In this scheme, a main pulse (pump pulse)
with maximum amplitude $a_0$ creates a high amplitude plasma wave and
collides with a secondary pulse of lower maximum amplitude $a_1$. The
interference of the two beams creates a ponderomotive beatwave pattern with
phase velocity $v_{bw}=0$, and thus enables to preaccelerate
background electrons.  Upon interacting with this field pattern, some background electrons gain enough momentum to be
trapped in the main plasma wave and then accelerated to high
energies. The force associated with this ponderomotive beatwave scales as $F_{bw}=2 k_0
a_0 a_1$, where $k_0$ is the central wavenumber of both pulses,
$F_{bw}$ is many times greater than the ponderomotive force
associated with the pump laser $F_{\text{pond}} \approx k_p a_0^2 $
since in an underdense plasma $ k_0>> k_p$. Therefore, the
mechanism is still efficient even for modest values of $a_0$ and
$a_1$. As the overlapping of the lasers is short in time, the
electrons are injected in a very short distance and can be accelerated to an almost monoenergetic beam. This concept has been recently validated in an experiment \cite{NatureCPI}, using two counter-propagating pulses. Each pulse had a duration of 30 fs at full width half maximum (FWHM), with
$a_0=1.3$, $a_1=0.4$. They were propagated in a plasma with electron density $n_e=7 \times 10^{18}
cm^{-3}$ corresponding to $\gamma_p=k_0/k_p=15$. It was shown that
the collision of the two lasers could lead to the generation of
stable quasi-monoenergetic electron beams. The beam energy could be
tuned by changing the collision position in the plasma.

The precise understanding of these experiments, as well as the
optimization of this process, motivate the present study. We have
used 1D Particle in cell (PIC) simulations to model electron
injection in the plasma wave at the collision of the two lasers,
and their subsequent acceleration. The PIC simulations are
compared to existing models \cite{Esarey} and show major
differences, such as the plasma fields behavior and the amount of
injected charge.

We first describe the fluid model developped in Ref.
\cite{Esarey,Fubiani}. In the linear approximation the wakefield
potential (due to charge separation) is a superposition of three
potentials \cite{Gorbunov} : $\Phi= \Phi_0 + \Phi_1 + \Phi_b$.
$\Phi_{0,1}$ are the charge separation potentials driven by the
laser pulses ${\bf a_{0,1}}$, $\Phi_0$ representing the main
accelerating structure we want to inject electrons in, and
$\Phi_b$ is driven by the beatwave. The normalized expressions of
the wakefields $\phi_i= e \Phi/mc$ are given by :
\begin{eqnarray}
\left(\frac{\partial^2}{\partial \xi_{0,1}^2} + k_p^2\right)\phi_{0,1} = &\frac{k_p^2}{2} <{\bf a_{0,1}}^2> \\
\left(\frac{\partial^2}{\partial t^2} + \omega_p^2\right)\phi_{b} = &\omega_p^2 <{\bf a_0. a_1}>
\label{eqbeat}
\end{eqnarray}
where $\xi_{0,1}=z\mp v_p t$ stands for the phase relative to the
pump and injection lasers, $k_p$ is the plasma wavevector and brackets $<.>$ denote the time-average over the fast varying scale ($1/(k_0c)$). The term $<{\bf a_0. a_1}>$
is the beatwave, appearing only during the collision of the two
lasers. Its spatial scale is given by $2 k_0$, for example for
circularly polarized lasers we have ${\bf a_0.a_1} = a_0(r,\xi_0)
a_1(r,\xi_1) \cos(2k_0 z)$.

The first approximation of this analytical model consists of neglecting the last term, $\phi_b$. This is supported by the fact that the density variation linked with this electrostatic potential scales as $\delta n_b /n = 4 k_0^2/k_p^2 \phi_b$ with $k_0/k_p>>1$. This density variation being limited to $\delta n_b /n \backsimeq 1$ even in a non-linear regime, the potential can not be greater than $\phi_b \backsimeq k_p^2/4 k_0^2$ which is often negligible compared to the other terms scaling as $\phi_{0,1} \backsimeq a_{0,1}^2$.

The second assumption is that we can separate the dynamics of the particles inside and outside the beatwave because the timescales are different (i.e. an electron will see a constant wakefield during an oscillation in a beatwave bucket).
An underlying hypothesis here is that there are two different species of electrons, those maintaining the wakefield, or fluid electrons, and those being trapped in the beatwave, or test electrons.

Starting from these hypothesis, one can even build an analytical
model when the lasers have modest  intensities ($a_0<1$) and are
circularly polarized. In that case, the beatwave pattern is time
independent. Using this analytical model, one can find an
analytical threshold for trapping (\cite{Esarey,Fubiani}) :
$2\sqrt{a_0 a_1} \geq u_z(\xi_{0min})$ where $u_z(\xi_{0min})$ is the
minimum normalized longitudinal momentum $p_z/mc$ for which the
electrons can be trapped in the wakefield. The experiments
described in ref\cite{NatureCPI} operated well above this
threshold.

A numerical implementation of this model is a particle tracking
code \cite{Fubiani} where test particles are pushed in prescribed
fields. The plasma fields are given by
$\mathbf{E_{pl}}=-\nabla(\Phi_0+\Phi_1)$ and the laser fields are
solutions of the paraxial wave equation with a linear plasma
response (gaussian beams). As expected, this model shows good
agreement with the analytical model in the circular polarization
case and allows to extend the scheme to the linear polarization
case for which analytical theory is untractable. For linear
polarization, the beatwave has a fast varying time dependence,
which leads to stochastic effects \cite{stoc}. In that case, the
trapping thresholds are even lower.

The main results of this model, that we will later on refer to as the "prescribed fields" model, are that electrons can be injected with modest values of $a_0$ and $a_1$ (linear regime). The resulting bunch is quasi-monoenergetic because all the electrons are injected in a short distance (where the beating occurs). Moreover, the charge is expected to be up to some hundreds of pC in the linear regime. 1D PIC simulations, where fields are linked to the motion of electrons, have also already been carried out \cite{kotaki}. They confirmed that an electron beam with low energy spread, low emittance and short bunch length can be obtained with modest values of $a_0$ and $a_1$.

Here we compare 1D PIC simulations with a 1D prescribed field
model. In this later model, we have used the 1D nonlinear theory of wakefield
generation to be able to compare the results in a more nonlinear
regime ($1<a_0<2$). The corresponding equation writes \cite{Esarey2}:
\begin{equation}
\label{eqrelf} \dif{^2\phi_{0,1}}{\xi_{0,1}^2} = k_p^2
\gamma_p^2\left[\beta_p\left(1-\frac{(1+<{\bf a_{0,1}}^2>)}{\gamma_p^2(1+\phi_{0,1})^2}\right)^{-1/2}
-1 \right]
\end{equation}

\begin{figure}[!ht]
\includegraphics[width=9cm]{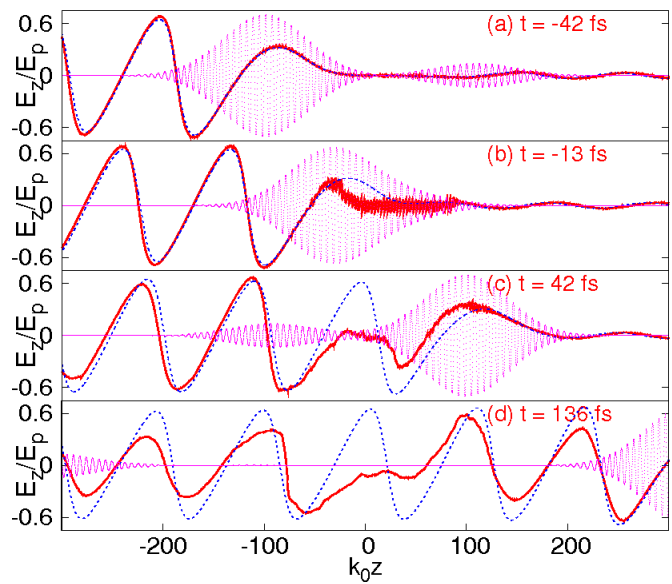}
\caption{\label{fig:wake} Longitudinal electric field computed at different times in PIC simulation (thick solid line), and in prescribed fields simulation (dotted line).}
\end{figure}
1D PIC simulations were carried out with the code CALDER
\cite{Erik}. For all the simulation results presented here, we
have used a simulation box of 6400 cells, each cell measuring
$0.1/k_0$. Both lasers were gaussian pulses with $30 fs$
duration at FWHM and wavelength $\lambda_0=0.8 \mu\mbox{m}$. To
avoid edge effects, the main pulse entered the plasma through a
density ramp of $100\mu\mbox{m}$, the maximum density being
$7\times10^{18} \mbox{cm}^{-3}$. The simulation box was kept fixed to
let the laser pulses enter the simulation correctly and we
otherwise used a moving window to follow the main pulse over long propagation distances..

Fig. \ref{fig:wake} shows the time evolution of the longitudinal
electric field in 1D PIC simulations when the polarizations of the
two pulses are parallel (solid line), compared to the fields
externally specified by equation (\ref{eqrelf}) (dotted line), for
$a_0=2$ and $a_1=0.4$. The electric field is given in linear wave
breaking field unit $E_p=c m_e \omega_p/e$ corresponding here to
$E_p= 250 \mbox{GV/m}$. The pump pulse propagates from left to
right. We have also plotted the transverse electric field (thin
dotted line) to show the position of the laser pulses. The
collision occurs at $z=0$ at time $t=0$. Fig.\ref{fig:wake}.a
shows the electric field 40 fs before the collision, it fits very
well with the solution of equation (\ref{eqrelf}). During the
collision of the two beams (Fig.\ref{fig:wake}.b) we clearly see a
small spatial scale pattern at $2 k_0$, created by the beatwave.
However, in this non-linear regime ($\delta n_b /n \backsimeq 1$)
we do not observe a superposition of this pattern with the usual
wakefields as is the case in ref \cite{Gorbunov}. The most
remarkable feature is the strong distorsion of the plasma wave at the position where the two pulses collide. This distorsion
remains after a long time (Fig.\ref{fig:wake}.c,
\ref{fig:wake}.d) and a numerical estimate shows that
the plasma wave amplitude is decreased by a factor of 10 at this position.

\begin{figure}[!ht]
\includegraphics[width=8.5cm]{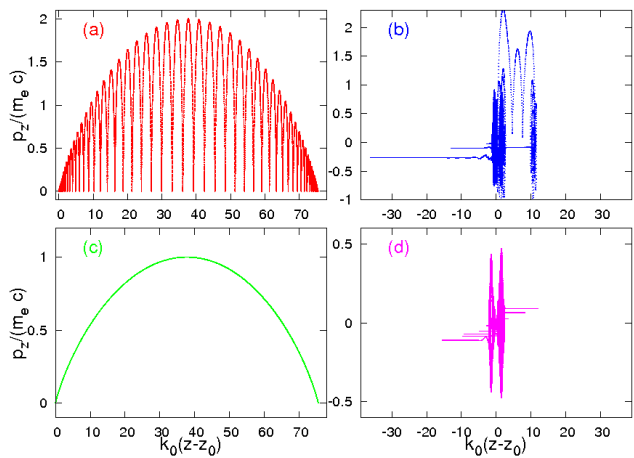}
\caption{\label{fig:phase} Typical trajectories of electrons in
phase space. a): Pump laser only, linear polarization ($a_0=2$);
b): Pump and injection lasers, parallel linear polarizations
($a_0=2$ and $a_1=0.4$). c): Pump laser only, circular
polarization ($a_0=2/\sqrt{2}$); d): Pump and injection lasers, circular
polarizations ($a_0=2/\sqrt{2}$ and $a_1=0.4/\sqrt{2}$)}
\end{figure}
This decrease of the wakefield amplitude follows from the fact that the electrons experiencing the collision of the lasers are trapped in the beatwave buckets and hence do not respond coherently (fluid-like) to the ponderomotive force of the main pulse. Without this fluid hypothesis, equation (\ref{eqrelf}) fails.  It would still hold if only some electrons were trapped, but here, as the phase velocity of the beatwave is zero, all electrons are trapped and kinetic effects cannot be treated as a perturbation.

To better understand this plasma wave "damping", we have performed
simulations where test electrons are submitted to the laser fields in
vacuum. In particular, this allows to understand electron motion
in the ponderomotive beatwave. Results of these simulations are
plotted in figure \ref{fig:phase} where we represent the orbits of
test electrons in $(z-z_0,p_z)$ phase space. Electrons are
initially taken at rest, their initial position $z_0$ is randomly
distributed around the collision position. Fig. \ref{fig:phase}.a
illustrates the trajectories of electrons experiencing the pump
laser field only (with $a_0=2$) : in the laser reference frame,
every electron has the same motion because they are pushed by the
same ponderomotive force. This fluid motion leads to the charge
separation which drives the electrostatic field known as the
plasma wave. On the contrary, when a
parallel polarized injection beam is included, trajectories are
drastically modified (Fig.\ref{fig:phase}.b). Electrons
are locally trapped in beatwave buckets, there is no large scale charge
separation and hence the plasma wake is no longer excited. This wake inhibition phenomenon
also occurs for circularly polarized lasers as shown in
Figs.\ref{fig:phase}.c and \ref{fig:phase}.d. This demonstrates
that the plasma wave distorsion is not related to the stochastic motion
of electrons, but really to their trapping in the beatwave
buckets.

\begin{figure}[!ht]
\includegraphics[width=8.5cm]{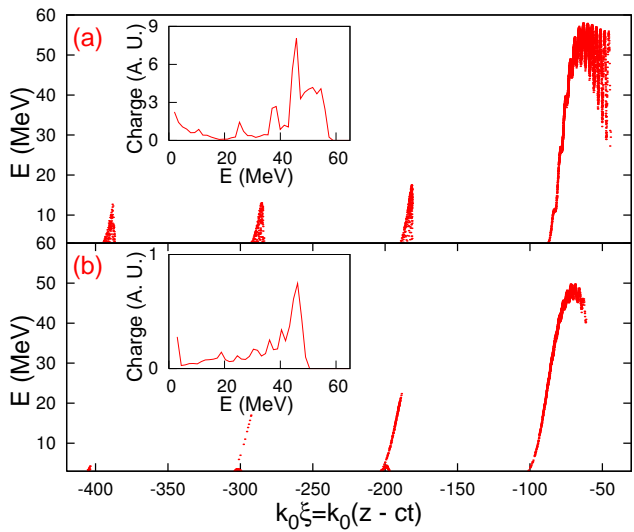}
\caption{\label{finalplot} Energy gain after $300 \mu$m as a function of relative phase $\xi = z-ct$, ($\xi=0$ corresponding to the maximum of the pump pulse). Top : prescribed fields simulation, bottom : PIC simulation. }
\end{figure}

We now investigate the consequences of this plasma wake inhibition on
the features of the electron beam. Figure \ref{finalplot} shows
the energy of the electron beam versus their phase with respect to
the laser (main plots) as well as the corresponding spectra
(inserts) $300\; \mu\mbox{m}$ after the collision of the two
beams. Fig.\ref{finalplot}.a corresponds to the prescribed field
simulation and Fig.\ref{finalplot}.b corresponds to the 1D PIC
simulation.
Qualitative differences on the energy spectra are only minor, both
spectra showing a peaked distribution around 45 MeV. Electrons are
trapped mainly in the first wakefield bucket, and less than ten
percent are trapped in the next ones. However, depending on the model, the trapped charge
differs by almost one order of magnitude.

In the prescribed fields model, electrons are preaccelerated in
the beatwave. As their initial velocity is lower than the wakefield phase velocity, they slip backward in phase. They are then trapped in the wakefield provided the energy gained in the beatwave was sufficient. In that case, electrons catch up with the plasma wave and are accelerated to high energies.
In the more realistic PIC simulation, the beatwave preaccelerates
electrons in the same way but the wakefield is inhibited and most
electrons just slip back in phase space without being trapped. In
order to be trapped, electrons need to be preaccelerated at the
end of the collision. As they slip back slower in phase space than
fluid electrons, they witness a restored wakefield suitable for trapping and acceleration when they reach the back of the pump pulse.
Therefore, the plasma wake inhibition reduces the phase space volume of injected particles. This leads to lower trapped charge and to a lesser extent, to a smaller energy spread in the PIC simulation.

\begin{figure}[!ht]
\includegraphics[width=8cm]{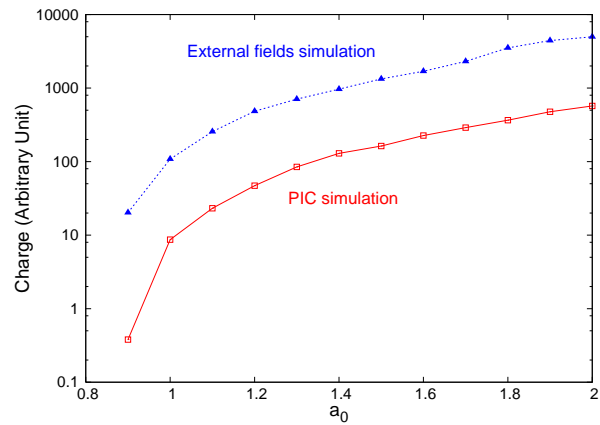}
\caption{\label{fig:comp} Comparison of injected charge obtained
in PIC simulations and in prescribed fields simulations.}
\end{figure}

Figure \ref{fig:comp} shows a more complete comparison of the charge
obtained in both simulations (logarithmic scale), with $a_0$
varying beetween $a_0=0.9$ and $a_0=2$. Here, $a_0=0.9$ corresponds to the trapping threshold. The hollow squares represent the
charge obtained in 1D PIC simulations ; the triangles represent
the charge obtained using the prescribed fields simulations. 
As we can see, the prescribed fields simulation overestimates the charge by a constant factor 8. The plasma wake inhibition and the reduction of the trapped charge
has also been confirmed in 3D hybrid PIC simulations.
\cite{agustin}.

In conclusion, although the previous modelling of the collision of
two laser pulses in an underdense plasma gives a good
understanding of the general processes at play in the injection of
electrons, we have observed and explained an important physical
process neglected to date. Using 1D PIC simulations, we have shown
that the beatwave does not only preaccelerate electrons but also
reduces the wakefield amplitude. In this process, the trapped
charge decreases by an order of magnitude and to a lesser extent,
the energy spread can be improved. Further studies will focus on
minimizing this effect and increasing the injected charge.

\bibliographystyle{aip}

\end{document}